# The News Says, the Bot Says: How Immigrants and Locals Differ in Chatbot-Facilitated News Reading

Yongle Zhang
College of Information, University of Maryland
College Park, Maryland, USA
yongle@umd.edu

Phuong-Anh Nguyen-Le
College of Information, University of Maryland
College Park, Maryland, USA
nlpa@umd.edu

Kriti Singh
College of Information, University of Maryland
College Park, Maryland, USA
kritisgh@umd.edu

Ge Gao
College of Information, University of Maryland
College Park, Maryland, USA
gegaoumd.edu

## ABSTRACT

News reading helps individuals stay informed about events and developments in society. Local residents and new immigrants often approach the same news differently, prompting the question of how technology, such as LLM-powered chatbots, can best enhance a reader-oriented news experience. The current paper presents an empirical study involving 144 participants from three groups in Virginia, United States: local residents born and raised there (N=48), Chinese immigrants (N=48), and Vietnamese immigrants (N=48). All participants read local housing news with the assistance of the Copilot chatbot. We collected data on each participant's Q&A interactions with the chatbot, along with their takeaways from news reading. While engaging with the news content, participants in both immigrant groups asked the chatbot fewer analytical questions than the local group. They also demonstrated a greater tendency to rely on the chatbot when formulating practical takeaways. These findings offer insights into technology design that aims to serve diverse news readers.

## 1 INTRODUCTION

News reading in today's digital era is transitioning from the reporter-oriented approach to a reader-oriented one. This transition opens exciting opportunities, as well as challenges, for researchers and system designers who aim to create tailored news experiences for readers. Recent research in information seeking, in the context of news reading and beyond, has advocated for the design of different supports for individuals from various social groups [4, 6, 18, 55, 113, 115]. Two groups central to this discussion are new immigrants (hereinafter referred to as "immigrants") and local residents (hereinafter referred to as "locals") living in the same society. Members of one group often differ from those of the other in terms of their most urgent information needs [10, 43, 59, 60], background knowledge about the local socio-informational environment [24, 34, 65, 74], and proficiency in the majority language used by most local media and essential services [11, 69, 80, 118]. These differences and how they interact can create differences in news consumption for immigrant versus local news readers.

In the current work, we leverage the Q&As between news readers, whether immigrants or locals, and an LLM-powered chatbot to explore directions for enhancing people's news reading experience. The chatbot serves a unique "dual role" in this research: it enables us to pinpoint a person's primary needs and interests in the news content, as that information is disclosed in the questions posed to the chatbot; meanwhile, it also provides us with the opportunity to consider chatbot design as one concrete approach to facilitating a reader-oriented experience. While a small number of existing projects have envisioned designs for news consumption along the latter direction, they usually adopted a staged setup. That is, news readers were tasked with imagining their chatbot usage in hypothetical scenarios (e.g., [51, 92]), or they seek the chatbot's assistance by asking prescripted questions (e.g., [145]).

The rest of this paper details our rationale, method, data analysis, and results from a between-subject study involving 144 individuals. Part of them were immigrants who relocated to the state of Virginia in the United States from Mainland China (N=48) and Vietnam (N=48), respectively. The rest of participants were locals (N=48) raised and living in the same region. All participants read through a shared set of news articles with the assistance of the Microsoft Copilot[1], an LLM-powered chatbot with state-of-the-art capabilities to process and answer reader-initiated questions through

[1]Microsoft Copilot is a chatbot developed by Microsoft and launched in February, 2023. It is a GPT4-based digital assistant designed to answer users' questions in a conversational format. It can perform searches and provide information based on the content of the current webpage or external public online sources.

Q&As. We performed a series of in-depth analyses of the questions initiated by immigrant news readers from both groups, as well as by locals. We also gathered multifaceted measures to assess the potential impact of news reading on participants' lives, including participants' self-rated value of their news reading experience and their detailed takeaways from the news as demonstrated in a retelling task.

Our work contributes to future research at the intersection of reader-oriented news, marginalized groups (e.g., immigrants), and chatbot design in the following ways:

- It provides a fine-grained typology of reader-initiated questions that span across 7 distinct categories and serve 4 high-level purposes, establishing a knowledge base to ground follow-up research on news reading;
- It demonstrates critical contrasts among the three news reader groups, offering quantitative evidence about how immigrants and locals differ in performing analytical thinking with the news content and in seeking practical guidance from it;
- It introduces novel methods, such as the retelling task and related framing analysis, to the study of chatbot-assisted news reading, enabling a precise understanding of the takeaways that people generate and the information sources that they draw from (e.g., content in the news article vs. the chatbot's responses);
- It outlines directions for future system design to better support diverse news readers, with particular attention to the needs and constraints of individuals in marginalized groups.

## 2 RELATED WORK

In this section, we synthesized three lines of prior work at the intersection of news informatics, HCI, and journalism. We began with literature indicating that reader-initiated questions are effective tools for understanding what people value in a given news article. We then summarized existing research that leverages Q&A chatbots to facilitate information seeking in news reading and beyond. Inspired by the above body of work, we wondered whether individuals from different social groups, such as immigrants and locals of the same region, would benefit equally from a news reading experience facilitated by today's Q&A chatbots. To explore this inquiry, we reviewed previous studies that highlight the unique needs and challenges of immigrants' news reading practices, compared to those of locals. We concluded with two research questions (RQs) guiding the current work.

### 2.1 Reader-Initiated Questions as the Indicator of What People Value

Recent discussions in news informatics and journalism emphasize the importance of supporting a reader's own needs and interests as they navigate through the reported content [81, 82, 99, 110, 123, 146]. However, given that news is often presented via a lean medium (e.g., text-based) and in a static manner (e.g., reading only), it is not easy for reporters to predict what their readers will care about before making the news content public [99, 102, 125]; nor can they effectively discern how readers will perceive the informativeness or usefulness of the news content while reading it [7, 48, 124].

One possible way to pinpoint the readers' needs and interests in the reported content is to examine the questions they ask about that content. Previous research has documented various ways to collect these reader-initiated questions. In the early days, writing letters to newspaper editors [134], participating in call-in radio programs [9], and filling out call-out surveys distributed by news organizations [49, 127] were popular means for readers to "have their say" and get their values heard. Over time, as the dissemination and consumption of information both go digital, online commenting has become another dominant means for readers to voice their thoughts and questions related to a news article [30, 97, 103].

While the above ways of question collection have proven valuable for creating a reader-oriented news experience, they share two significant constraints. One constraint is the time delay in readers receiving responses to what they have asked. As previous research pointed out, it is challenging for news producers to give immediate reactions to readers' questions, although they often refer to those questions when drafting subsequent news pieces on relevant topics [32, 67, 117]. The other constraint considers the lack of scalability. In particular, some of the traditional means, such as call-in radio programs, were not designed to collect questions from a large number of news readers. Filling out surveys or leaving online comments potentially allows more readers to share thoughts about the reported content; however, active participants in these activities usually feature specific attributes and cover only a small sector of the entire reader population [12, 49, 58].

### 2.2 Information Seeking Through Q&As with Chatbots

Chatbots have the potential to respond to user-generated questions promptly and on a large scale. Their success in assisting people with information seeking has been demonstrated in various scenarios, including the collection of neighborhood and community information by local residents [1, 44, 56, 96], the understanding of learning materials by students [22, 109, 138], the explanation of medical documents for patients [15, 63], or the examination of consent forms for research participants [139]. Across these scenarios, chatbots not only provide their users with the requested information, but also enable other stakeholders - such as community workers, teachers, doctors, and researchers - to recognize the information seeker's primary interests, as revealed through the questions being asked. When it comes to news reading, chatbots, in their traditional design, have primarily been implemented as recommendation tools, pushing news notifications to users following a greeting message (e.g., [36, 57, 143]). They rarely handle open Q&A initiated by news readers, despite operating with a conversational interface.

A small body of recent work has explored the concept of chatbots offering Q&A assistance during a readers' real-time navigation through news articles. Much of this work does not study the news reader's actual use of chatbots; instead, it focuses on envisioning the design of conversational agents for news consumption in general [51, 53, 92]. In particular, Nordberg and Guribye conducted co-speculation workshops with individuals experienced in HCI, journalism, and other relevant fields. Participants brainstormed scenarios in which chatbots, as well as other conversational agents, might facilitate their daily news consumption. Data obtained from these workshops highlighted critical design guidelines for effective conversational agent design [92]. In a more recent project, Hoque et al. asked Mturkers to write down questions that they had after a news reading task. Their analysis indicated that participants tended to highlight different questions to ask, depending on the belief about who would process and respond to those questions (e.g., a Q&A chatbot or a human journalist) [51]. Several other projects in this space have invited human participants to work with functioning chatbots for news reading. These chatbots, whether commercial systems or in-house prototypes, were all developed with certain capabilities to answer reader-initiated questions. That said, for the purpose of initial testing, participants were usually instructed to have Q&As with the chatbot using a preset list of questions (e.g., [19, 21, 145]).

To enhance the chatbot's capability of handling a wide range of reader-initiated questions, ongoing research in NLP and information retrieval has been making advancements in the text processing step. This endeavor has resulted in various formats of conversational agents, ranging from in-house systems trained with domain-specific news corpora (e.g., [66, 68, 79, 106, 133]) to LLM-powered Q&A chatbots for information seeking in general.

## 2.3 News Reading by Different Social Groups: Immigrants vs. Locals

As the potential of Q&A chatbots gets increasingly unlocked, we believe it is time to consider not only *whether* chatbots can promote reader-oriented news consumption, but also *how* they can provide suitable support for news readers from different social groups. For the interest of the current work, we pay attention to the contrast between immigrants and locals living within the same society. Previous research has evidenced contrasts in how these different groups attend to local news, as well as in how they generate takeaways from the news. That said, many of these conclusions are descriptive in nature. None of them have considered the emerging context of news reading that involves a chatbot assistant.

*2.3.1 Attending to local news.* In cross-disciplinary literature on information seeking, immigrants and locals are frequently depicted as having separate goals to prioritize during daily news reading and beyond [6, 18, 55, 73, 93, 94, 113, 115]. For example, Oh and Butler conducted online surveys and follow-up interviews to compare the information needs of new immigrants and domestic residents relocating to a new city in the United States. Immigrants reported a significantly greater demand for information related to their survival needs, whereas domestic people exhibited an inherent familiarity with the local socio-informational infrastructure [93]. Caidi et al. synthesized literature on immigrants' social adaptation in Canada and elsewhere. They outlined a trajectory in which an immigrant's needs may gradually evolve from seeking settlement-related help to collecting more advanced information (e.g., tacit social knowledge for civic participation), requiring considerable support for making these transitions. In contrast, local adults often need much less help in accessing information at more basic levels [18]. Many empirical studies have also indicated that, compared to locals, immigrants often need additional scaffolding in seeking domain-specific information of their interests, such as understanding health insurance policies [104, 132, 147], getting their questions about employment and job opportunities answered [78, 101, 113, 144], and interpreting online reviews of local services [3, 89, 120].

Besides holding various needs, immigrants and locals can also encounter distinct challenges in processing a given piece of news information [40, 80, 83]. In a recent project, for instance, Gao et al. studied how Chinese immigrants in Japan and the United States consumed COVID-relevant news during the global pandemic. They found that immigrants seldom engaged with the mainstream media of the given society, despite having a strong willingness to understand the local COVID situation [40]. During interviews in this research and a few others, immigrants shared that they often had questions about how to interpret the news content. Adding to it, there was a lack of channels for them to discuss their questions with others and obtain answers [18]. Limited English fluency often emerged as another roadblock, further complicating the information seeking efforts and effectiveness of immigrants compared to locals [23, 34, 39, 47, 83, 84, 113]. Unfortunately, most news informatics research to date has not considered tailoring the design of news reading tools for individuals from various social groups.

Our current research takes Q&A chatbots as a point of departure to investigate potential differences between immigrants and locals in terms of their demand during news reading and, hence, for reader-oriented assistive tools. In this context, Q&A chatbots are positioned both as our research device and target design space to explore. Specifically, we leverage Microsoft Copilot, an LLM-powered chatbot to assist immigrant and local participants with news reading tasks. As the chatbot responds to reader-initiated questions in real-time and throughout the task process, it compiles a dataset of questions asked by participants. This dataset will enable us to compare the primary needs and interests of news reading between the two target groups, following the rationale implied by the literature reviewed earlier. We pose the following RQ:

***RQ1****: What questions will immigrants and locals raise, respectively, during their news reading with the assistance of a chatbot? Will they seek different types of assistance from the chatbot?*

*2.3.2 Generating takeaways from local news.* Another equally noticeable contrast between immigrants and locals, if not more, lies in how people derive takeaways from news reading. Previous research on news consumption has suggested some potential differences in this regard (e.g., [5, 28, 50, 61, 69, 72, 114, 140, 142]). For instance, Yang et al. [140] surveyed 84 Chinese internationals living in the Midwestern United States. These participants self-reported a strong focus on using local media to learn about societal norms and to acquire English vocabulary for language learning. Similarly, Alencar and Deuze [5] conducted focus groups with three immigrant groups in the Netherlands and Spain. Participants shared that they derived insights from local news to aid their adaptation to the local society. However, they were unsure whether their interpretations were of comparable quality to those of the locals.

Notably, while the above literature offers essential knowledge about the takeaways immigrants draw from news reading, it lacks comprehensive details. The types of information that constitute their takeaways remain unspecified. Concrete evidence on how immigrants gather information from various accessible sources to formulate their takeaways has also been lacking. Also, the contrasts between immigrant and local news readers were often inferred from studies that focused on a single group, rather than from direct comparisons between groups.

Our current research contributes to the existing literature by directly comparing the takeaways generated by immigrants and locals from news reading. Inspired by prior work on the broad field of text comprehension (e.g., [45, 70, 85]), we developed a news retelling task (see Section 3.2 for details) to better document the information that a news reader includes in their takeaways. This approach enables us to better quantify the differences between news reader groups regarding their focus, as well as the information source they rely on (e.g., the news article itself vs. the chatbot's responses). We address the following RQ:

***RQ2****: What information will immigrants and locals focus on, respectively, in the takeaways they draw from news reading? Will they pull information from different sources to formulate their takeaways?*

## 3 METHODS

To answer the aforementioned RQs, we designed a between-subject study involving 144 participants living in the same region. This study received IRB approval from our institution and was evaluated as posing minimal risks to human participants. Before beginning the study, participants were provided with an informed consent form outlining the purpose of the research, the types of data being collected, the procedure, and the compensation (i.e., 25 U.S. dollars per person). The form also explained how their information would be stored and anonymized. Additionally, participants were informed of their right to withdraw from the study at any time without penalty and were assured that participation was entirely voluntary. They were encouraged to contact the research team with any questions before signing up for the study.

All study sessions were conducted remotely, where participants used their own computers to perform the task and stay connected with a researcher via the screen-sharing mode of Zoom. We adopted this setup to ensure that participants were fully engaged in the news reading task rather than multitasking during the remote session. The researcher was instructed to avoid constantly monitoring the participant's screen share, minimizing the influence of their presence on the participant's task experience. Participants were tasked with reading a set of news articles about local housing in the region where they live, a topic frequently highlighted as being relevant and important to everyone's daily life [41, 80, 93, 116]. Each participant had one hour to read through the entire set of news. An LLM-powered chatbot, Microsoft Copilot, was provided to assist participants in their news reading process, offering the information they requested in a Q&A format. After that, participants filled out a questionnaire to report their overall news reading experiences and then worked on a news retelling task. Below, we elaborate on each aspect of our research design in further detail.

### 3.1 Participants

A total of 144 individuals from three social groups participated in our study: two groups of immigrants and one group of locals. All immigrants were individuals who relocated from overseas to live and work in Virginia, United States. We recruited them by sharing social media posts and distributing paper flyers among the immigrant communities of Virginia. All immigrant participants rated themselves as having limited working proficiency in speaking the majority language of the United States (i.e., English).

One group of these immigrants was originally from China (N = 48). Among them, 31 participants self-identified as female, and 17 self-identified as male. Their average age was 26.94 years (S.D. = 5.00). Their average length of residence in the United States was 5.34 years (S.D. = 3.94). Their native language was Mandarin.

The other group of immigrants came from Vietnam (N = 48). Among them, 24 participants self-identified as female, and 24 self-identified as male. Their average age was 26.81 years (S.D. = 4.21). Their average length of residence in the United States was 6.31 years (S.D. = 4.23). Their native language was Vietnamese.

The local group consisted of individuals who were raised in Virginia and had lived there for most of their lives (N = 48). Among them, 19 self-identified as female, and 29 self-identified as male. Their average age was 35.33 years (S.D. = 10.13). They all identified themselves as native English speakers with full proficiency.

### 3.2 Task Procedure

All participants received task instructions in their native language and completed the study with the assistance of a researcher fluent in that language. We allocated a one-hour session to each participant. This duration was sufficient for most people to finish reading the entire set of articles, as confirmed by our pilot testing with individuals of varying reading speeds and English proficiency levels. All participants went through four successive phases during their participation:

*3.2.1 Preparation phase.* To ensure all participants had the same understanding of how Microsoft Copilot works, we conducted a tutorial for each person before they began the news reading phase. During the tutorial, participants were introduced to the basic functionalities of this chatbot, including information searching both within and beyond a given webpage, text processing in the format of summarizing and paraphrasing, translating text across languages, and synthesizing information per the user's requests. The content of this tutorial was identical for both local and immigrant groups. At the end of the tutorial, participants set up Microsoft Copilot on their own computer's browser and then completed a practice trial of news reading with the chatbot's assistance. This step allowed the researcher to recognize and address any usability issues that people might encounter during their interaction with the chatbot. As shown in the practice trial, none of our participants experienced confusion or technical challenges in using Copilot for news reading before starting the formal task.

*3.2.2 News reading phase.* Participants were tasked with reading a shared set of four news articles prepared by the research team. The order of news reading across the four articles was fully counterbalanced between individual sessions, resulting in 24 possible combinations. Within each social group, there were two individuals under each of these 24 combinations. All participants were instructed to formulate their questions, if any, in the language they preferred at the moment (e.g., English, Mandarin, or Vietnamese). They also had complete freedom to ask the chatbot any questions as they read through the news articles. There was *no* requirement forcing people to interact with the chatbot.

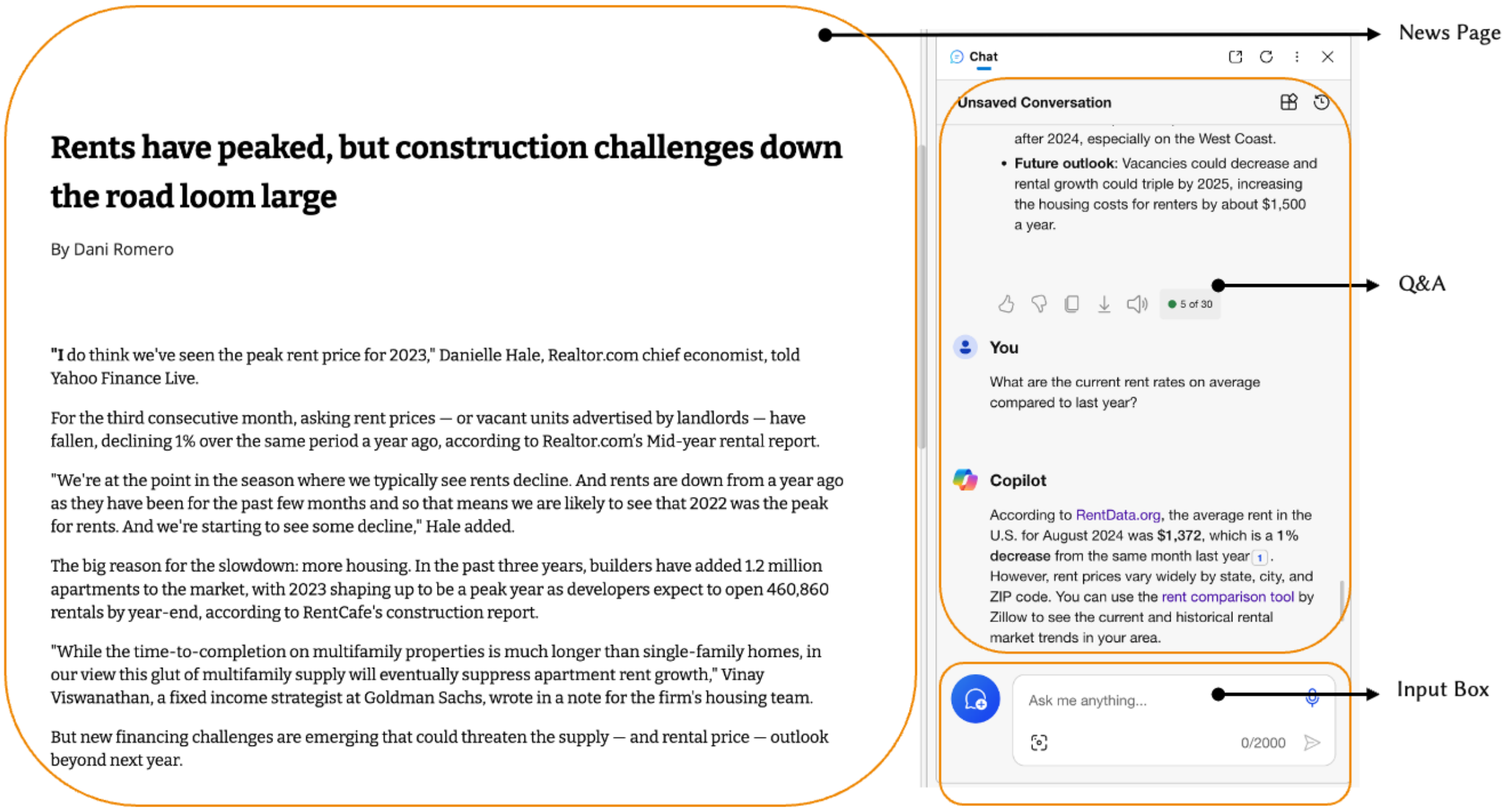

**Figure 1: An illustration of the task interface. The left panel displays the news page, which presents the news article along with its source media. The right panel shows the Q&A page along with the input box for participant questions.**

*3.2.3 Survey phrase.* Right after their reading of each news article, participants were asked to answer a short list of questions about their overall news experience.

*3.2.4 Reflection phrase.* After survey completion, participants were directed to a writing page. They were instructed to retell the news story they had just read, with the goal of informing another person who shares a similar background to themselves. To prevent directly copying and pasting from the news article or the history of their Q&A with the chatbot, we made both the article page and the chatbot page inaccessible once a participant moved to this retelling task. There were no restrictions on the length or the language to use for this retelling. After a participant completed all the above activities, we acknowledged their contribution to the research and provided compensation.

Looking back on the actual time participants spent in each phase of the formal task, they allocated an average of 5.5 minutes reading each news article, followed by 2.5 minutes responding to the survey. The average time spent on the retelling task was five minutes per news article. We did not observe significant differences in time allocation among participants in the three groups.

### 3.3 News Materials

There was an identical set of four news articles for all participants to read. These articles were selected from a larger collection of news from reputable local and national sources (e.g., The New York Times, Yahoo News, and Virginia Mercury), using search keywords related to real estate and residential rentals. To ensure the quality of the selected articles, we conducted a pilot evaluation on Prolific. Each Prolific worker was randomly assigned nine news articles to read and evaluate on a list of 7-point Likert scales. Their evaluations considered four distinct dimensions recommended in prior research [27, 42], including the news article's topic, length, style, and complexity:

- Topic: the selected news articles should discuss housing information that can be applied to individuals living in the state of Virginia, aligning with the general interest of our participants from this specific region;
- Length: the selected news articles should be of reasonable and comparable length, ideally around five minuses of reading time per article, allowing the reading of multiple articles and follow-up tasks to be completed within one hour.
- Style: the selected news articles should not be opinion-based;
- Complexity: the selected news articles should not be too complex for general readers to follow.

The four news articles used in our final task met all the above criteria, as confirmed by evaluations detailed in the Appendix. It was also confirmed by all participants that none of them had consumed these articles before attending our study.

## 3.4 Interfaces

Figure 1 demonstrates the task interface for news reading. We hosted the four news articles via a website set up by our research team. The interface replicated the common design of news webpages, ensuring a consistent layout as participants proceeded across articles. Certain visual elements, such as pictures and advertisement blocks, were excluded from our webpage design, which enabled us to avoid potential confounding factors that might affect the participants' news reading experience and our data analysis.

## 3.5 Measurements

We obtained data on the following aspects to answer our RQs:

*3.5.1 Personal background.* Before participants attended the study session, we asked each of them to report their self-identified *Attachment to the Topic of Housing* (1 = not at all, 7 = very much) and *Prior Experience with Chatbots* in daily work and life (1 = not at all, 7 = very much) in addition to other demographic information (i.e., *Age, Gender*). We considered these measures as control variables in later data analysis, as suggested by prior literature [25, 86].

*3.5.2 Reader-initiated questions.* We obtained the text log of each participant's Q&As with the chatbot. This data provided us with a lens to understand people's needs and interests, as reflected in the informational requests they sent to the chatbot in real time while reading the news. A total of 1,845 questions were identified from the log data, consisting of 699 questions asked by Chinese immigrants (with 316 questions asked in Mandarin), 557 questions asked by Vietnamese immigrants (with 316 questions asked in Vietnamese), and 589 questions asked by locals.

*3.5.3 Reader-generated takeaways.* After they read a news article, we collected the text content generated in each participant's news retelling. This data enabled us to grasp the concrete takeaways people had derived from their news reading. Our implementation of the retelling task was inspired by prior work on text comprehension, where participants were asked to write down what they remembered from reading a document [45, 70, 85]. By shifting the main focus of this task from individual memory and cognition to information sharing with an external audience, we made the task more suitable for capturing people's takeaways from news reading. These takeaways, as the literature on news consumption and journalism has suggested [16, 54, 90], typically involve people generating their interpretations of the news article rather than merely memorizing its content. Our participants generated a total of 576 retelling pieces: one piece per news article, and 4 pieces per person. Among them, 165 pieces were formulated in Mandarin, and 184 pieces were in Vietnamese.

*3.5.4 Overall News Reading Experiences.* We used survey questions to gain insights into each participant's overall news reading experience, even though they did not directly address our RQs. Part of these questions asked about the *Value* of news reading perceived by each person, derived from prior research on societal adaptation [5, 137] (e.g., "Reading this news enriched my knowledge about essential local services and affairs on the relevant topic," 1 = strongly disagree, 7 = strongly agree; Cronbach's $\alpha$ = .80). The rest of the questions asked about each person's *Workload* during news reading, adapted from the NASA-TLX scale [46] (e.g., "How much mental and perceptual activity was required in this news reading process?" 1 = not at all, 7 = very much; Cronbach's $\alpha$ = .75).

# 4 DATA ANALYSIS AND RESULTS

We performed three subsections of analyses to pinpoint the differences, as well as similarities, between immigrants and locals in their news reading assisted by a chatbot. These analyses focused on comparing participants' self-initiated questions during news reading (**RQ1**), self-generated takeaways after news reading (**RQ2**), and survey reports on their overall news reading experience.

In each subsection, we first introduce the results of our descriptive analysis. We then detail the comparisons between Chinese immigrants and locals, followed by corresponding comparisons between Vietnamese immigrants and locals. All comparisons used locals as the reference group.

## 4.1 Reader-Initiated Questions Asked by Immigrant vs. Local Participants (RQ1)

*4.1.1 Typologies and distributions.* We began the analysis of reader-initiated questions by identifying the types of questions exchanged between news readers and the chatbot. This step set the foundation for more detailed comparisons between immigrant and local news readers under **RQ1**.

Specifically, we manually coded the entire set of 1,845 questions obtained from all participants. To facilitate this process, we hired two human translators who were fluent bilinguals in Mandarin-English, as well as two translators fluent in Vietnamese-English, to generate the English translation of all the Q&A logs. The translation work allowed for follow-up analyses that involved researchers who did not speak some of these languages. The first and the last authors of the current paper independently reviewed all the Q&A logs. Each person worked on a subset of the data, identifying mutually exclusive categories that could best describe the nature of the questions initiated by our participants. Subsequently, they cross-checked the questions and categories generated by each other.

The above steps were repeated until the two researchers had examined all subsets of the data and, accordingly, revised all categories that had been generated. These categories and their definitions were provided to a third person, blind to all previous steps, for another round of independent categorization of all reader-initiated questions. We obtained a high inter-coder reliability score at this step (Cohen's Kappa = .92), indicating satisfactory robustness of the converged rules. After that, the two researchers and the aforementioned third person attended group coding sessions together, resolving remaining inconsistencies and finalizing the categorization of all questions.

Table 1 details the seven categories of questions identified through our manual coding process[2]. These categories are grouped under

[2]Instances where participants acknowledged successful receipt of a message from the chatbot (e.g., "Thank you.") were excluded from the table and further analysis. These acknowledgements were rare (8 cases in total) and less relevant to our RQs.

**Table 1: Questions asked by local and immigrant participants across categories.**

| Theme/Category | | Definition | Sample Question of Locals | Sample Question of Immigrants |
|---|---|---|---|---|
| **Literal Comprehension** | Word Explanation | Lexical explanations of words and phrases that appear in the given news | "What is the meaning of glut?" | "Slump 是什么意思？" [English translation: What does slump mean?]<br>"Deter nghĩa là gì?" [English translation: What does deter mean?] |
| | Text Summary | A concise representation of the selected text from the given news | "Can you briefly summarize this news?" | "请简单复述一下这篇新闻的内容。" [English translation: Please briefly summarize the content of this news.]<br>"Neu y chinh cua bai bao nay." [English translation: Present the main idea of this article.] |
| **Analytical Thinking** | Viewpoint Synthesis | An integration of insights and perspectives for explanations on a reader-specified issue conveyed in the news | "Why is housing growth slow?" [Context: The news article discusses various issues associated with this slow growth.] | "为什么文章里说邻近的田纳西和北卡罗来纳州比弗吉尼亚便宜？" [English translation: Why does this news article say that (the housing expenses in) Tennessee and North Carolina are cheaper than Virginia?] [Context: The news article discusses various issues that are associated with the expense difference across these states, among other content.]<br>"Từ bài báo, tại sao giá thuê nhà lại tăng?" [English translation: From the news article, why has rent increased?] [Context: The news article discusses the rise in rent prices, outlining multiple factors contributing to this increase.] |
| | Referential Fact | Factual information related to content mentioned in the news but requiring external sources for further details | "What is the year-to-year rent in Arlington, VA?" [Context: The news mentions yearly rent in Arlington but does not specify rates across years.] | "美国有哪些城市实行了租金管制？" [English translation: Which cities in the United States have implemented rent control?] [Context: The news article mentions rent control in the United States but does not specify which cities have implemented it.]<br>"Có chương trình của chính phủ để hỗ trợ việc thuê nhà không?" [English translation: Are there government programs for rental assistance?] [Context: The news article mentions a proposed rent control measure but does not detail any existing government programs.] |
| | Broader Impact | Societal effects or consequences related to the issues reported in the news but going beyond its explicit content | "How does Joe Biden's Renters Bill of Rights affect non-subsidized housing?" [Context: The news mentions the Renters Bill of Rights but does not discuss unsubsidized markets.] | "该文章对于中国新移民有什么样的影响？" [English translation: What impact does this news have on Chinese new immigrants?] [Context: The news article mentions proposed actions on rentals but does not discuss the implications for Chinese immigrant populations.]<br>"Luật mới này có lợi cho chủ nhà hay người mướn nhà?" [English translation: Is this new law more beneficial for homeowners or renters?] [Context: The news article discusses proposed actions regarding the current rental market but does not address specific stakeholders or the impact on these groups.] |
| **Practical Guidance** | | Recommendations or advice that encourage specific behaviors of the news reader but go beyond what this news has explicitly discusses | "I am a renter in VA, what should I be paying attention to and what advice can you give?" | "你能给我一些找到便宜公寓的建议吗？" [English translation: Can you give me advice on finding an affordable apartment?]<br>"Có thể làm sao để tiết kiệm được tiền thuê nhà?" [English translation: What to do in order to save money on rent?] |
| **Conversational Repair** | | The chatbot's notice of the need to rework its immediate previous response to the news reader | "That's not quite relevant to my question. Can you answer my question by referring to the news I am currently reading?" | "我问的不是存款的利率而是房屋贷款利率。" [English translation: I asked about mortgage interest rates, not deposit interest rates.]<br>"Ý tôi là Hoa Kỳ chứ không phải Mỹ Đình." [English translation: I mean Hoa Kỳ (United States), not Mỹ Đình (a place in Vietnam).] |

**Note:* Immigrant participants asked questions in both English and their native languages. In this table, we focus on the questions in their native languages for the language variety among all examples.

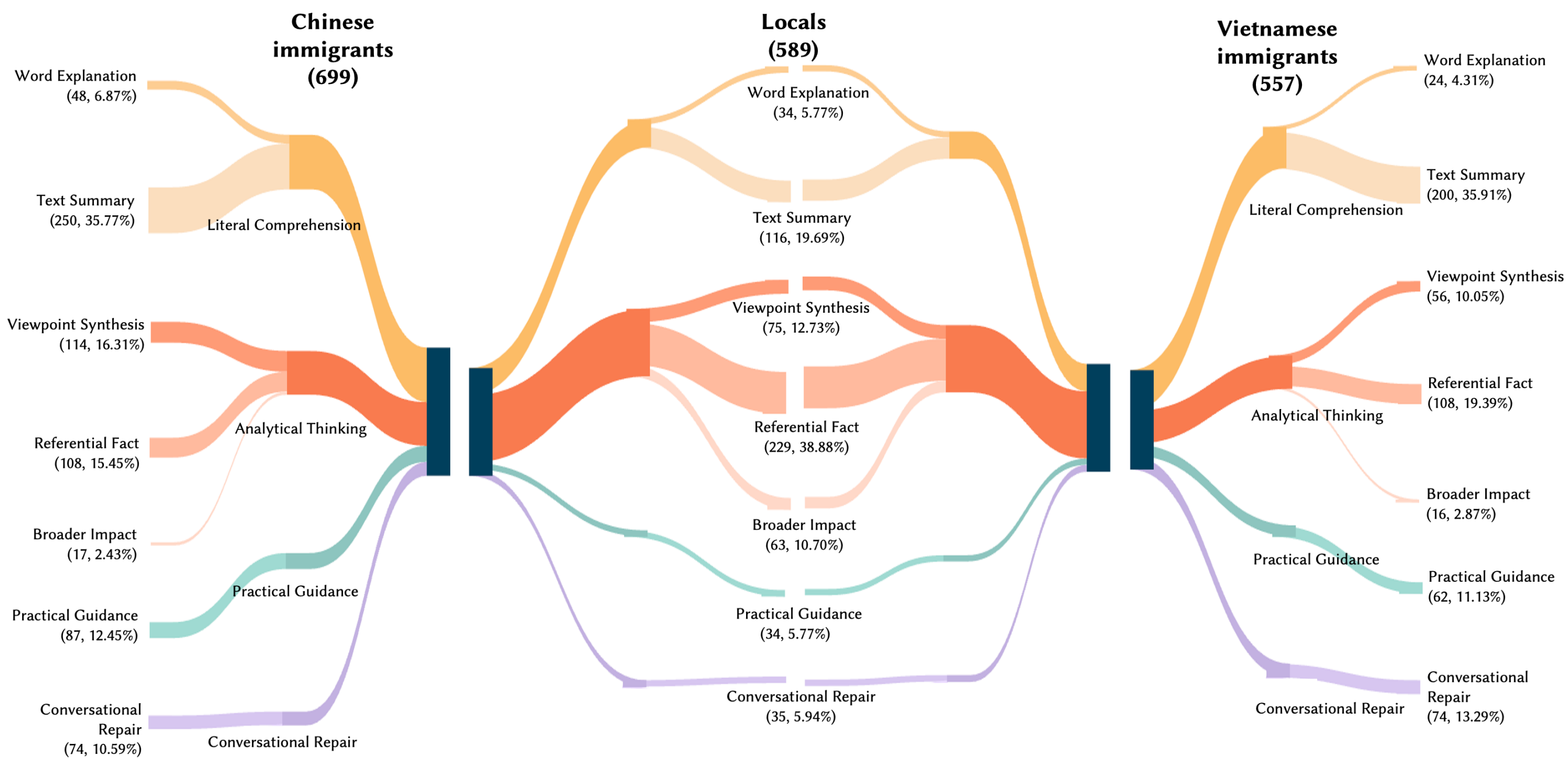

**Figure 2: Distribution of questions among immigrant and local participants. Left and right wings are the counts and percentages of questions initiated by immigrant groups. Middle wings are those of questions initiated by locals.**

four high-level themes. Three of these themes –*Literal Comprehension*, *Analytical Thinking*, and *Practical Guidance* –reflect participants' informational needs and interests during news reading. The fourth one –*Conversational Repair* –captures their coordination with the chatbot when the Q&A did not go smoothly.

Based on this categorization, we counted the raw number as well as the percentage of questions under each category. Figure 2 illustrates the distribution of these counts across the three participant groups. We observed potential differences between immigrants and locals in terms of the question categories they prioritized. In particular, compared to participants in the local group, individuals in both immigrant groups appeared to ask the chatbot more for assistance with *Literal Comprehension* and *Practical Guidance*, but less for *Analytical Thinking*. Next, we applied more rigorous analyses to verify the observed contrasts.

*4.1.2 Comparisons between participants in the Chinese immigrant group and the local group.* We performed a set of negative binomial regression analyses to assess the comparison between Chinese immigrants and locals in terms of their likelihood of asking the chatbot questions from each category. This analysis is particularly useful when the dependent variable consists of discrete counts and exhibits overdispersion, as is the case with our current data on the counts of reader-initiated questions.

We built one negative binomial regression model for each question category. In the model, the number of questions asked by a participant was used as the dependent variable. The main predictor was the participant group (i.e., Chinese immigrant or local). We included the following control variables across all models: the participant's *Age* (i.e., number of years), Gender (i.e., female or not), *Attachment* to the topic of housing (i.e., rating on a 7-point scale), *Prior Experience* with chatbots (i.e., rating on a 7-point scale), and the *Total Number of Questions* the person asked. Table 2 specified the effect of the participant group in predicting people's likelihood of asking the chatbot questions from each of the 7 categories, as well as indexes of the overall model fit.

In short, our results suggested that, with all control variables held constant, the expected counts of questions exchanged between Chinese immigrants and the chatbot were significantly fewer than those exchanged between locals and the chatbot in the categories of *Referential Fact* (decreased by 60%) and *Broader Impact* (decreased by 75%). Conversely, Chinese immigrants tended to exchange more questions in the categories of *Text Summary* (increased by 103%), *Practical Guidance* (increased by 93%), and *Conversational Repair* (increased by 126%). There was no significant difference between the two groups in other categories.

*4.1.3 Comparisons between participants in the Vietnamese immigrant group and the local group.* We conducted another set of analyses to assess the comparison between Vietnamese immigrants and locals. The model choice and variable setting were identical to those used for the comparison between Chinese immigrants and

**Table 2: Effect of participant group (i.e., Chinese immigrants vs. locals) on the questions asked per category.**

| | Participant Group (Chinese Immigrant vs. Local) | | | | | | Overall Model Fit | | | | |
|---|---|---|---|---|---|---|---|---|---|---|---|
| | B | S.E.(B) | df | *p* | Exp(B) | 95%CI | Log-likelihood | AIC | BIC | Deviance/df | Pearson $\chi^2$/df |
| Word Explanation | 0.08 | 0.47 | 1 | .87 | 1.08 | [0.43, 2.72] | -117.35 | 250.70 | 271.21 | 0.88 | 1.09 |
| *Text Summary | 0.71 | 0.15 | 1 | .00 | 2.03 | [1.52, 2.71] | -204.80 | 425.60 | 446.11 | 1.50 | 1.20 |
| Viewpoint Synthesis | 0.27 | 0.22 | 1 | .23 | 1.31 | [0.85, 2.04] | -157.82 | 331.64 | 352.16 | 1.18 | 1.06 |
| *Referential Fact | -0.92 | 0.18 | 1 | .00 | 0.40 | [0.28, 0.57] | -195.47 | 406.93 | 427.44 | 1.29 | 1.16 |
| *Broader Impact | -1.41 | 0.38 | 1 | .00 | 0.25 | [0.12, 0.51] | -100.86 | 217.73 | 238.24 | 0.97 | 1.16 |
| *Practical Guidance | 0.66 | 0.27 | 1 | .01 | 1.93 | [1.15, 3.24] | -122.91 | 261.82 | 282.34 | 1.26 | 1.29 |
| *Conversational Repair | 0.81 | 0.32 | 1 | .01 | 2.26 | [1.21, 4.22] | -131.85 | 279.70 | 300.22 | 1.14 | 1.06 |

*Note:* * indicates statistical significance.

**Table 3: Effect of participant group (i.e., Vietnamese immigrants vs. locals) on the questions asked per category.**

| | Participant Group (Vietnamese Immigrant vs. Local) | | | | | | Overall Model Fit | | | | |
|---|---|---|---|---|---|---|---|---|---|---|---|
| | B | S.E.(B) | df | *p* | Exp(B) | 95%CI | Log-likelihood | AIC | BIC | Deviance/df | Pearson $\chi^2$/df |
| Word Explanation | -0.66 | 0.44 | 1 | .13 | 0.52 | [0.22, 1.23] | -94.73 | 205.45 | 225.97 | 0.86 | 0.95 |
| *Text Summary | 0.49 | 0.15 | 1 | .00 | 1.63 | [1.21, 2.19] | -198.32 | 412.64 | 433.16 | 1.60 | 1.23 |
| *Viewpoint Synthesis | -0.45 | 0.22 | 1 | .04 | 0.64 | [0.41, 0.98] | -130.33 | 276.66 | 297.17 | 1.27 | 1.36 |
| *Referential Fact | -0.69 | 0.19 | 1 | .00 | 0.50 | [0.35, 0.74] | -192.12 | 400.23 | 420.75 | 1.25 | 1.06 |
| *Broader Impact | -1.62 | 0.37 | 1 | .00 | 0.20 | [0.10, 0.41] | -92.52 | 201.04 | 221.55 | 1.05 | 1.53 |
| *Practical Guidance | 0.90 | 0.28 | 1 | .00 | 2.45 | [1.42, 4.25] | -117.50 | 251.00 | 271.52 | 1.40 | 1.57 |
| *Conversational Repair | 0.56 | 0.27 | 1 | .04 | 1.76 | [1.04, 2.98] | -125.17 | 266.33 | 286.85 | 1.47 | 1.52 |

locals. Table 3 specified the effect of the participant group in predicting people's likelihood of asking the chatbot questions from each of the seven categories and the overall model fit.

With all control variables held constant, the expected counts of questions exchanged between Vietnamese immigrants and the chatbot were significantly fewer than those exchanged between locals and the chatbot in the categories of *Viewpoint Synthesis* (decreased by 36%), *Referential Fact* (decreased by 50%), and *Broader Impact* (decreased by 80%). That said, Vietnamese immigrants tended to exchange more questions with the chatbot in the categories of *Text Summary* (increased by 63%), *Practical Guidance* (increased by 145%), and *Conversational Repair* (increased by 76%). There was no significant difference between groups in other categories.

***To recap,*** our analyses revealed that immigrant and local news readers both raised questions within a shared set of categories. However, there were significant differences in terms of which question categories were more valued by whom. Compared to locals, individuals in both immigrant groups were less likely to ask questions that aligned with the high-level purpose of *Analytical Thinking*, and more likely to raise questions under the high-level themes of *Literal Comprehension* and *Practical Guidance.* Figure 3 illustrates these contrasts in summary, with locals set as the reference group across all the categories.

## 4.2 Takeaways Generated by Immigrant vs. Local Participants (RQ2)

*4.2.1 Framing elements and distributions.* We followed Entman's framing theory to examine all retelling pieces obtained from data collection, considering the retelling content as a reflection of the takeaways generated by news readers [33]. This theory is widely referenced in prior research on news media and journalism (e.g., [29, 111, 126]), usually for analyzing the reporter's narrative about a news event. We adopted it to analyze how our participants framed their takeaways from news reading, allowing for structured comparisons between immigrant and local news readers under RQ2.

In his framing theory, Entman identified four primary elements of a narrative: descriptive statement, causal attribution, moral evaluation, and treatment recommendation. We applied these four concepts to understand the framing of all 576 retelling pieces in our data. Our review of this retelling data followed a similar workflow to our categorization of the Q&A data. That is, we began by translating all the retelling pieces into English. After that, we iterated between independent coding and group discussions. By the end of this process, we reached a high level of consensus regarding how Entman's concepts could be used to understand the content of our participants' retelling pieces. We asked two independent coders, blind to all previous steps, to identify the primary elements of each retelling piece with the instructions we provided. They reached high inter-coder reliability (Cohen's Kappa = .94) and resolved the remaining inconsistencies through further discussions.

Table 4 demonstrated the three primary elements confirmed and applied to our data analysis, within the context of one retelling piece generated by our participant. Specifically, we adopted Entman's original concepts of *Descriptive Statement* and *Causal Attribution*, as the same two elements of content also frequently appeared in our retelling data. We proposed a new concept termed *Action Suggestion*, as our participants often provide "suggestions to others" when retelling the news. However, our participants, as news readers, rarely made explicit moral evaluation or treatment recommendations in their retelling, which differs from how news reporters would do in their narratives.

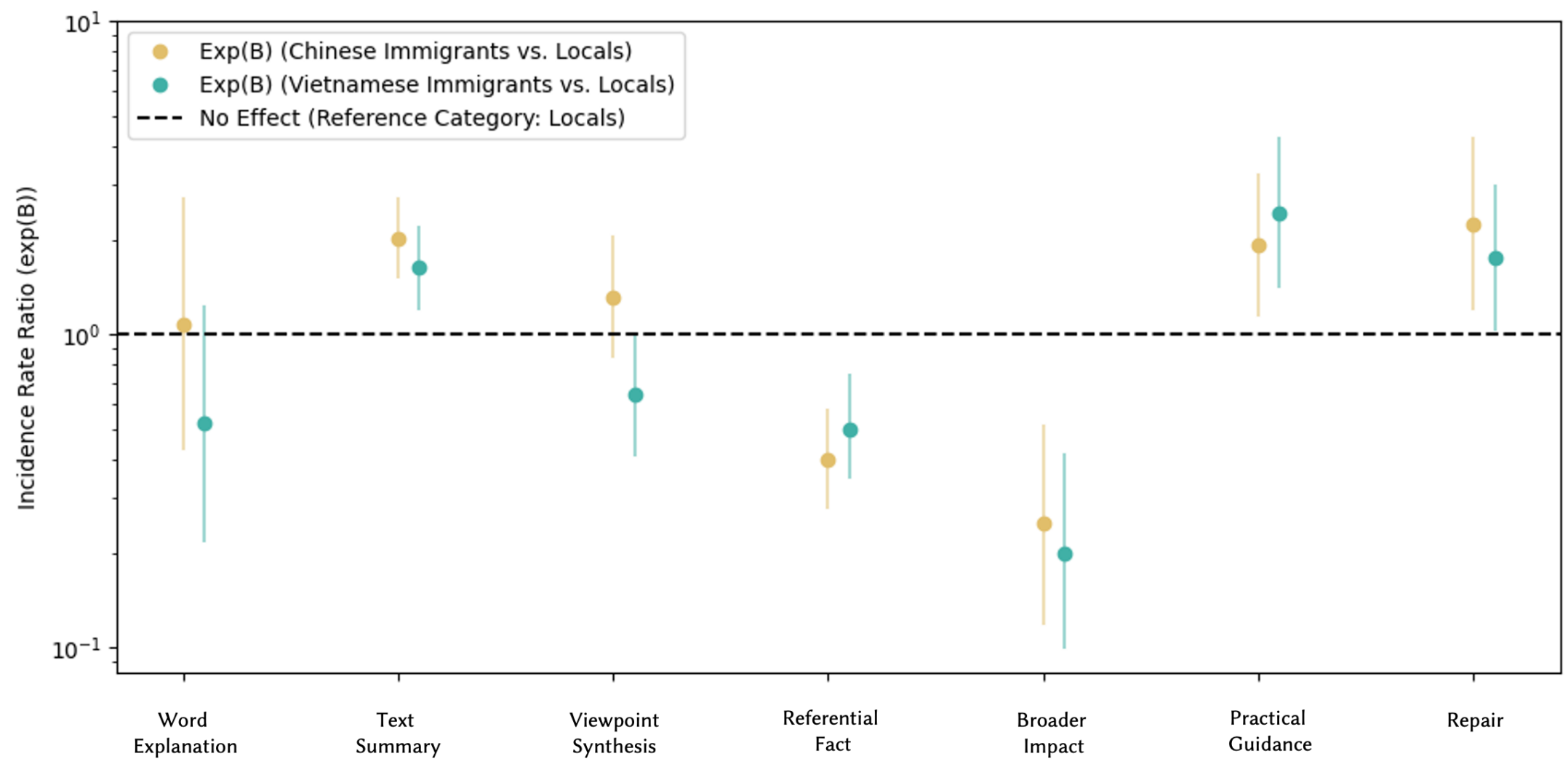

**Figure 3: Exponentiated coefficients (exp(B)) from all seven negative binomial regression analyses that examined the effect of participant group (i.e., Chinese Immigrants or Vietnamese Immigrants vs. locals) on the questions asked per category.**

**Table 4: Retelling example and its identified segment under each framing element.**

| **Example of One Retelling Piece** | | |
|---|---|---|
| "Hey, how's it going? We recently talked about the housing market in Virginia, and I just read something interesting that explains the rental side of things. Apparently, rent prices have been going down for a few months. Now might be a good time to lock in a lease. Rent rates vary across the country, depending on the metro area. The recent drop in prices is linked to the number of apartments being built. Recently, millions of units have been constructed across the country, driving down overall rent prices." | | |
| **Framing Element** | **Definition** | **Identified Segment** |
| Descriptive Statement | Description of the situation, condition, or context presented in the news. | Rent prices have been going down for a few months.<br>Rent rates vary across the country, depending on the metro area.<br>Millions of apartment units have been built recently. |
| Causal Attribution | Identification and explanation of the causes or reasons behind an issue or event, answering the questions of what leads to what and why. | The recent drop in prices is due to the large number of apartments being built recently. |
| Action Suggestion | Recommendations, proposed solutions, or suggested actions in response to the issues presented in the news or relevant to readers' daily lives. | Now is a good time to try to lock up a lease. |

By deconstructing the content of each retelling piece according to the three primary elements, we were able to trace the source of the segment under each element. Specifically, we compared every segment of a retelling piece against two bodies of text: the text from the news article that participants had read, or the text from the chatbot's responses that participants had received during news reading. This comparison resulted in three possible sources for every segment of a retelling piece:

- *Article:* the content in this segment overlaps exclusively with information provided in the news article;
- *Chatbot:* the content in this segment overlaps with information provided by the chatbot;
- *External Insight:* neither the article nor the chatbot provided information that overlaps with the content in this segment.

Figure 4 illustrates the distribution of participants who had leveraged each possible source to formulate their retelling content under each element. Overall, these counts indicate that local participants cited content from the *Article* in their retelling segments of *Descriptive Statement* and *Causal Attribution* more than immigrants did. In contrast, immigrants cited content from the *Chatbot*'s responses in their retelling segments of *Action Suggestion* more than locals did. Next, we applied more rigorous analyses to detail the contrasts in retelling between immigrants and locals.

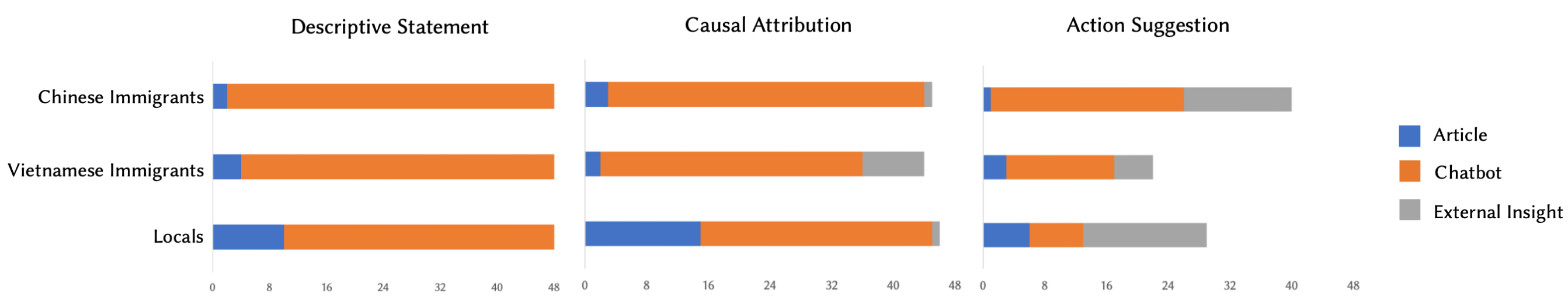

**Figure 4: Distribution of the use of information sources by participant group (i.e., Chinese immigrants, Vietnamese immigrants, or locals).**

*4.2.2 Comparisons between participants in the Chinese immigrant group and the local group.* We conducted a set of binary logistic regression analyses to compare Chinese immigrants and locals in terms of their likelihood of including each of the three framing elements in their retelling pieces. The dependent variable of each regression model considered whether a framing element was included in this person's retelling (i.e., absent or present). The primary predictor was the participant group[3] (i.e., Chinese immigrant or local). We also included the same control variables as those used in models under Section 4.1. Our results showed that, compared to participants in the local group, those in the Chinese immigrant group were significantly more likely to include *Action Suggestions* in their retelling (increased by 259%). This was the only framing element that demonstrated a significant difference between them.

We then proceeded to the comparisons of sources. Specifically, among participants who included segments of *Descriptive Statement* in their retelling pieces, we set up three binary logistic regression models to predict whether those segments had used *Article* (i.e., no or yes), *Chatbot* (i.e., no or yes), or *External Insight* (i.e., no or yes) as its information source, respectively. The same steps were repeated for the segments of *Causal Attribution* and then for *Action Suggestions*. These models all included the participant group (i.e., Chinese immigrant or local) as the main predictor, along with the same set of control variables used previously. Table 5 specified the effect of participant group in predicting the source of each element of retelling content, as well as indexes of the model fit.

Our results suggested that, compared to locals, Chinese immigrants were more likely to rely on *Chatbot*'s responses as the source for generating *Causal Attribution* (increased by 2,260%) and *Action Suggestion* (increased by 834%) relevant takeaways in their news retelling. They were less likely to rely on news *Articles* as the source for generating *Causal Attribution* relevant takeaways (decreased by 95%). Other aspects of the retelling analyses did not reveal significant differences between Chinese immigrants and locals.

*4.2.3 Comparisons between participants in the Vietnamese immigrant group and the local group.* We performed a similar set of binary logistic regression analyses to draw the comparison between Vietnamese immigrants and locals. The model choice and variable setting were identical to those used in Section 4.2.2. The only change was that the main predictor, participant group, had been redefined (i.e., Vietnamese immigrant or local). We found no significant differences between Vietnamese immigrants and locals in terms of their likelihood of including any of the three framing elements in their retelling pieces. That said, these two groups leveraged the source differently. Table 6 specified the effect of participant group in predicting the source of each element in the retelling content, as well as indexes of the model fit.

Compared to locals, Vietnamese immigrants were more likely to rely on *Chatbot*'s responses as the source to generate *Action Suggestion* relevant takeaways in their news retelling (increased by 4,969 %). They were less likely to rely on the news *Article* as the source to generate *Causal Attribution* relevant takeaways (decreased by 95%). They were also less likely to use *External Insights* - information that overlaps with neither the article content nor the chatbot's responses - to formulate *Action Suggestion*-relevant takeaways (decreased by 94%). Other aspects of the retelling analyses did not reveal significant differences between Vietnamese immigrants and locals.

***To recap***, our analyses revealed that immigrant and local news readers demonstrated different tendencies in relying on each potential source to generate takeaways from news reading. Compared to locals, individuals in both immigrant groups were more likely to refer to *Chatbot*'s responses when formulating *Action Suggestions*-relevant takeaways. Also, both were less likely to refer to the news *Article* when generating takeaways related to *Causal Attribution*. There were several differences between the two immigrant groups in terms of how they compared to the locals, but the overall results demonstrated more of their similarities. Figure 5 illustrates these contrasts in summary, with locals set as the reference group across all categories.

[3]For all analyses in Sections 4.2.2 and 4.2.3, we used each participant as the unit of analysis, rather than each retelling piece written by the participant or each framing element within the retelling. In our dataset, specifically, although each participant created one retelling piece for every news article they read, the four retelling pieces by the same individual included an identical set of framing elements in our dataset. Similarly, the same individual appeared to rely on the same set of sources when generating descriptive statements, causal attributions, and action suggestions in their retellings. These data patterns indicated a high degree of intrapersonal consistency both in people's framing of the retelling and in their source use for the retelling, reflecting each person's individual style.

**Table 5: Likelihood of source occurrences across framing elements by Chinese immigrants vs. locals.**

| | Participant Group (Chinese Immigrant vs. Local) | | | | | | Overall Model Fit | | |
|---|---|---|---|---|---|---|---|---|---|
| | B | S.E. | Wald $\chi^2$ | $p$ | Exp(B) | 95%CI | $\chi^2$ | Nagelkerke $R^2$ | N |
| **Article** | | | | | | | | | |
| Descriptive Statement | -1.50 | 0.97 | 2.39 | .12 | 0.22 | [0.03, 1.49] | 24.15 | 0.42 | 96 |
| *Causal Attribution | -3.02 | 0.94 | 10.25 | .00 | 0.05 | [0.01, 0.31] | 31.34 | 0.46 | 91 |
| Action Suggestion | -2.28 | 1.39 | 2.69 | .10 | 0.10 | [0.01, 1.56] | 14.45 | 0.39 | 69 |
| **Chatbot** | | | | | | | | | |
| Descriptive Statement | 1.50 | 0.97 | 2.39 | .12 | 4.46 | [0.67, 29.65] | 24.15 | 0.42 | 96 |
| *Causal Attribution | 3.16 | 0.95 | 11.08 | .00 | 23.60 | [3.67, 151.80] | 36.52 | 0.51 | 91 |
| *Action Suggestion | 2.23 | 0.74 | 9.11 | .00 | 9.34 | [2.19, 39.87] | 14.04 | 0.25 | 69 |

**Table 6: Likelihood of source occurrences across framing elements by Vietnamese immigrants vs. locals**

| | Participant Group (Vietnamese Immigrant vs. Local) | | | | | | Overall Model Fit | | |
|---|---|---|---|---|---|---|---|---|---|
| | B | S.E. | Wald $\chi^2$ | $p$ | Exp(B) | 95% CI | $\chi^2$ | Nagelkerke $R^2$ | N |
| **Article** | | | | | | | | | |
| Descriptive Statement | -1.37 | 0.84 | 2.67 | .10 | 0.25 | [0.05, 1.32] | 19.28 | 0.32 | 96 |
| *Causal Attribution | -2.95 | 0.95 | 9.66 | .00 | 0.05 | [0.01, 0.34] | 25.01 | 0.39 | 90 |
| Action Suggestion | -0.06 | 1.35 | 0.00 | .96 | 0.94 | [0.07, 13.33] | 11.45 | 0.33 | 51 |
| **Chatbot** | | | | | | | | | |
| Descriptive Statement | 1.37 | 0.84 | 2.67 | .10 | 3.93 | [0.76, 20.32] | 19.28 | 0.32 | 96 |
| Causal Attribution | 1.00 | 0.62 | 2.64 | .11 | 2.72 | [0.81, 9.08] | 10.86 | 0.16 | 90 |
| *Action Suggestion | 3.93 | 1.26 | 9.74 | .00 | 50.69 | [4.31, 596.85] | 21.49 | 0.46 | 51 |

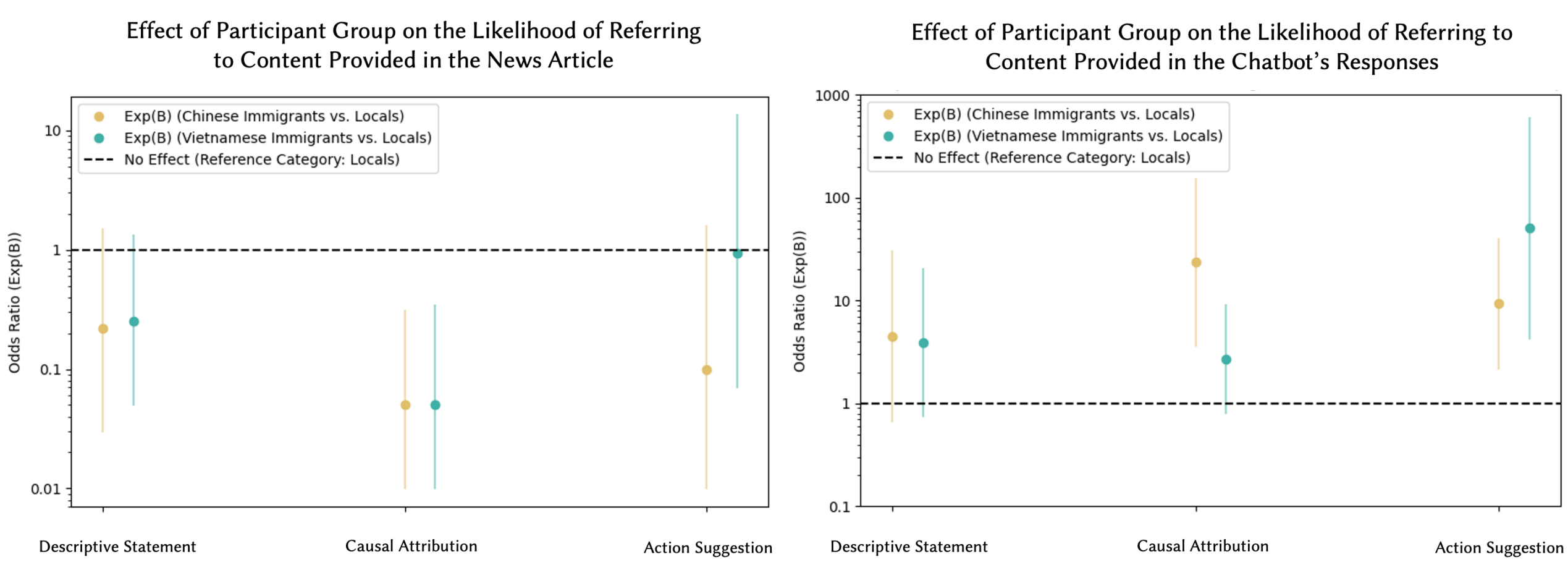

**Figure 5: Exponentiated coefficients (Exp(B)) from binary logistic regression analyses that examined the effect of participant group (i.e., Chinese Immigrants, Vietnamese Immigrants, or locals) on the likelihood of citing *Article* (left) and *Chatbot* (right) provided content in takeaways related to *Descriptive Statement, Causal Attribution, and Action Suggestion.***

## 4.3 Overall News Experience Perceived by Immigrant vs. Local Participants

We performed two ANOVAs to compare the overall *Value* perceived from the chatbot-facilitated news reading among participants in the three groups, as well as the overall *Workload* they experienced. Before conducting these analyses, we checked the normality of the dependent variables and assessed the homogeneity of variances across groups, confirming that the assumptions of ANOVA were met. Although these analyses did not directly relate to our **RQ1** and **RQ2**, they provided insights that contextualize other results.

In both ANOVA models, we selected the participant group (i.e., Chinese immigrants, Vietnamese immigrants, or locals) as the main independent variable. We included the same set of control variables as in other analyses, including a person's *Age, Gender*, *Attachment* to the topic of housing, *Prior Experience* with chatbots, and the *Total Number of Questions* they had exchanged with the chatbot. The dependent variable was participants' self-rated experience of news reading, collected via Likert scales and, thereby, treated as continuous measures.

Our results indicated a significant main effect of the participant group on a person's perceived *Value* of news reading: $F(2, 136) = 3.71$, $p < .05$. In particular, both Chinese immigrants (M = 5.31, S.D. = 0.66) and Vietnamese immigrants (M = 5.28, S.D. = 0.99) perceived greater *Value* from reading the given news articles than the locals did (M = 4.86, S.D. = 0.92). There was no significant difference in the *Workload* perceived by Chinese immigrants (M = 3.43, S.D. = 0.90), Vietnamese immigrants (M = 3.26, S.D. = 1.01) and locals (M = 3.31, S.D. = 0.83): $F(2,136) = 0.75$, $p = .48$.

# 5 DISCUSSION

To wrap up, we conducted a between-subject study to compare chatbot-facilitated news reading among three groups: Chinese immigrants, Vietnamese immigrants, and locals living in the same region. Our results revealed a list of statistically significant and conceptually meaningful contrasts between the immigrant groups and the local group. In the rest of this section, we reflect on three key aspects of our work: the reader-chatbot interaction as a viable means of assessing news readers' needs (Section 5.1), the varying depths of sensemaking between immigrant and local news readers (Section 5.2), and the chatbot's role in influencing immigrants' takeaways from news reading (Section 5.3). We outline relevant design insights at the end of each subsection.

## 5.1 Reader-Chatbot Interaction as a Viable Means of Assessing News Readers' Needs

To the best of our knowledge, the current research is among the first behavioral studies that leverage reader-chatbot interaction to pinpoint news readers' needs. This approach complements prior work in important ways.

In particular, research in journalism and social media has long been interested in referring to news readers' needs to enhance tailored news production and delivery [8, 31, 129, 136]. Participants in those studies typically made self-reports via surveys and interviews outside of their news reading process (e.g., [13, 14, 35, 37, 76, 119, 141]). More recent work, including many by HCI scholars, has utilized eye tracking and other computer-based techniques to capture an individual's interests as they process the news content (e.g., [17, 77, 121]). While these behavioral tracking techniques provide valuable information on how readers distribute visual attention across a news article, they often come with challenges in interpretation.

Exchanges with the chatbot can externalize a news reader's interests *during* their news reading process, contributing to recent initiatives that aim to promote news readers' interactive engagement with the news content [123]. Such an approach helps uncover the nuanced needs of news readers with relatively low effort. More importantly, as our results have demonstrated, the reader-chatbot interaction will be particularly valuable in helping immigrant news readers voice their needs and questions in addition to supporting local news readers. The former group has often been placed on the periphery of decision making regarding what to feature in a given society's mainstream news, whose perspectives to include in the news, and where to gather feedback for the next iteration of news production and delivery.

## 5.2 Varying Depths of Sensemaking Between Immigrant and Local News Readers

Our analyses of reader-initiated questions provide typologies and benchmarks for follow-up work that aimed at creating reader-centered news experiences. Question categories identified in the current work reflect three layers of needs and interests from the perspective of news readers: the literal layer (i.e., *Word Explanation* and *Text Summary*), the analytical layer (i.e., *Viewpoint Synthesis*, *Referential Fact*, and *Broader Impact*), and the practical layer (i.e., *Practical Guidance*). They formulate a "ladder of sensemaking," priming individuals to comprehend the news at varying levels of depths.

With the landscape of questions appearing consistent across the three news reader groups, we observed clear contrasts between immigrants and locals regarding their likelihood of asking questions under each specific category. The raw counts and percentages reveal an overall trend: immigrants were more likely to leverage the Q&A chatbot for *Literal Comprehension* of the given news, and locals tended to utilize the same chatbot more for *Analytical Thinking* of the news content. This difference was further unpacked in detail through our regression analyses. Specifically, participants of the local group prioritized the collection of *Referential Fact* that can supplement and, sometimes, verify information already described in the given news article. They also made strategic use of the chatbot to learn about *Broader Impact* of the reported content, exploring connections between various societal issues beyond the scope of the given news article. These behaviors enable the individual to not only understand but also critically reflect on or assess the information being presented [98].

In contrast, both groups of our immigrant news readers were more likely to seek information regarding *Text Summary* and *Practical Guidance.* They also tended to experience more difficulties in having their questions understood by the chatbot, as evidenced by their requests for *Conversational Repair.* Some of these results, such as the demand for summaries and repairs, were not entirely surprising given the language profile of the immigrant individuals in our sample. Especially for immigrants who speak a minority language of the society, achieving a satisfactory level of literal comprehension of the news can often be challenging [2].

The issue warranting the most attention is the abrupt transition of both immigrant groups from achieving *Literal Comprehension* of the news to seeking *Practical Guidance* from it. The entire sector of *Analytical Thinking* questions appeared significantly less prominent in the Q&A data of these participants compared to that of locals. Such a contrast might suggest that the news content, if unverified or misleading, would expose immigrants to greater risks. Immigrants are likely to remain unaware of this risk, given that they perceived the value of their news reading to be high (and it was significantly higher than the rating of locals). This lack of depth in immigrants' news reading highlights a "cold start" problem, where individuals may have insufficient background knowledge to guide them in performing further steps [75]. The more a news interface –whether it includes chatbot assistance or not –counts on the readers themselves to seek and interpret relevant information, the more it may widen the sensemaking gap between immigrant and local news readers.

Future chatbot design should consider having the chatbot provide proactive support to news readers, rather than only passively reacting to reader-initiated questions. This support is best provided after a news reader has initiated some self-exploration with the news content, thereby preventing the loss of human agency, as Sundar discussed in his recent piece [122]. One possibility is to have the chatbot synthesize questions typically asked by locals, such as those considering *Referential Fact* and *Broader Impact*, and use them to scaffold and prompt immigrant news readers' sensemaking at the analytical level. There can be added value in extending such support to local readers as well, ensuring a diverse range of questions flowing in both directions - from locals to immigrants and vice versa. This reciprocal exchange of perspectives could enrich the sensemaking experience for all, rather than tethering the information needs and preferences of one group (e.g., locals) to those of another (e.g., immigrants). The question categories and corpus obtained from our research, along with our analyses, provide essential materials for design explorations in this direction.

## 5.3 Chatbot's Role in Influencing Immigrants' Takeaways from News Reading

We employed novel methods (i.e., the retelling task) to investigate the takeaways people derived from news reading and the sources they relied upon. Compared to the local group, both immigrant groups appeared to rely more on the chatbot's responses when formulating takeaways in terms of the *Action Suggestion* shared with others. They generated *Causal Attribution* relevant insights to interpret the news event they had read, even though these interpretations were less likely to be directly derived from the content of the news article. We concluded from these results that, especially for immigrant news readers, the chatbot had functioned as an important "proxy" connecting the news with its audiences.

The results highlighted above align with recent research on the broad topic of human-AI interaction. For instance, it has been found across various scenarios that individuals with insufficient knowledge in a task domain tend to rely heavily on AI's outputs for task completion in that domain, especially when they strongly believe in the AI's capability to process complex data (e.g., [52, 62, 108, 135]). In our research context, immigrant news readers possess less societal knowledge compared to local readers. Such disparity may lead to the former's greater reliance on chatbot responses when forming their news takeaways.

Furthermore, we notice interesting differences between the two immigrant groups in terms of the source they relied on for drawing takeaways from news reading. In particular, Vietnamese immigrants were less likely than locals to count on *External Insights* when framing *Action Suggestions* -relevant information in their takeaways. The similar contrast, however, was not found in the comparisons between Chinese immigrants and locals. While it is difficult to provide a definitive explanation, we suspect that some variations in communal practices between Vietnamese and Chinese immigrants, as well-documented in social science literature, have partially contributed to our results. In brief, Chinese immigrants living in the same society are frequently described as maintaining strong intragroup connections to preserve their shared identity [20, 87]. In contrast, studies with Vietnamese immigrant groups in the United States have revealed greater subgroup differences rooted in mixed historical and religious identities, which hinders the formation of a cohesive community [71, 112, 131]. In Park et al.'s recent survey study, for example, participants in Vietnamese groups reported significantly lower levels of ingroup cohesion and engagement compared to other ethnic groups living in the United States [95]. This cultural attribute may lead Vietnamese immigrants to avoid disclosing certain information, such as their own insights and views on the news content, when retelling the news to others with a similar background.

In envisioning future chatbot design for reader-centered news experience, our work points to a dilemma that deserves attention, particularly for immigrant news readers. On the one hand, chatbots hold the potential to direct news readers to information that they were likely to overlook otherwise (as discussed in Section 5.2); on the other hand, news readers could become dependent on the chatbot's responses, potentially diverting their focus away from the original news content. In this latter case, interacting with chatbots would not help enrich news readers' understanding but merely shift their attention from one set of information to another.

To mitigate the above issue, our suggestion is not to prioritize the news article's content over the chatbot's responses, *nor* vice versa. These sources may each have its inherent strengths and biases, and the influence can vary depending on the reader group's characteristics and situational demands. In particular, the increasing advancement of LLMs reinforces their potential to assist immigrants in news reading through a convenient and interactive interface. Compared to traditional systems, conversational interfaces enable news readers to express their information needs more naturally [100, 130]. The further integration of LLMs allows chatbots to navigate large-scale information and generate responses tailored to readers' queries. However, recent studies have highlighted that LLMs may carry inherent biases stemming from their training data [38, 107]. News articles generated by journalists often adhere to established quality standards, such as prioritizing official administrative sources [91, 128]. That said, these standards do not completely eliminate biases, as individual journalists and news organizations may incorporate their own perspectives into news content production [91, 105]. These trade-offs between different information sources underscore the necessity for news readers to be aware of and critically engage with a diverse pool of information. Future news reading tools and interfaces should explicitly encourage news readers to consider diverse sources, especially when drawing takeaways to inform their actions in daily life.

## 6 LIMITATION AND FUTURE WORK

While we carefully reflected on our methodological choices before implementing them, several limitations of this work remain. We believe it is important to share further reflections on them.

Our research sample included three groups of participants: Chinese immigrants, Vietnamese immigrants, and local residents. We focused on these populations based on a joint consideration of several factors. From a communication perspective, while text-based conversations in Mandarin and Vietnamese are both supported to some extent by today's natural language processing techniques, Vietnamese is often considered a lower-resourced language compared to Mandarin. Including immigrants from both groups allowed us to add linguistic diversity as a crucial dimension to the investigation of immigrant news readers' needs [26, 64, 88, 144]. Additionally, both immigrant groups are well represented among the immigrant population in the geographic area where our study took place, facilitating smooth recruitment and data collection. That said, our sample did not encompass the full range of immigrant groups or individuals from various regions. Future research could examine chatbot-assisted news reading by individuals with more diverse backgrounds, using the typologies and analyses detailed in our work as benchmarks for comparison.

Regarding the demographic information collected from each participant, we asked participants to report their educational level (i.e., whether they had obtained any degrees from higher education) and their work status (i.e., whether they were working adults or students). However, both pieces of information were not measured in categorical formats. For example, we did not ask for the majors in which participants earned their degrees or the specific occupations they held. As all participants identified themselves as holding higher education degrees and being working adults at the time of the study, we excluded these two constant variables from data analyses. Additionally, our measure of language proficiency did not specifically isolate reading or writing ability but instead captured a combination of both as a whole. Follow-up studies should collect more fine-grained and possibly a wider range of demographic measures than we did, exploring the relationship between these variables and individuals' news reading behaviors.

Moreover, all the news articles used in our study focused on the topic of housing, covering both residential rentals and real estate. We selected this topic because prior research on immigrants' information-seeking behavior has identified housing as a key focus for immigrants [18, 113, 120]. In the context of Virginia, data from the Department of Housing and Urban Development indicates that both immigrant and U.S.-born individuals in the area allocate at least twenty percent of their income to housing costs. Background responses collected during our study aligned with this data, suggesting that all participants felt a personal connection to the topic of housing, regardless of residential status. While this information confirmed housing as an appropriate news reading topic for our research purposes, it limited our ability to explore participants' news reading behaviors on other topics.

Last but not least, it is important to acknowledge that, by conducting a controlled experiment and adopting a quantitative approach, we missed opportunities to examine more natural news reading behaviors for better ecological validity. Some specific setups of our study, such as requiring screen sharing during remote task sessions, may have inevitably introduced issues of social desirability in our data collection, despite our efforts to minimize such effects through established protocols. We encourage future studies to further investigate chatbot-assisted news reading with a wider range of participant groups, scenarios, and methods. We also advocate for follow-up research involving other key stakeholders, such as professional journalists and news media policymakers, to explore the promises and challenges of news reading in today's AI-powered era. The research materials and results generated from our study pave the way for this exciting area of exploration.

## 7 CONCLUSION

We investigated the differences in information needs between immigrant and local individuals during chatbot-assisted news reading, as well as the takeaways they formulated from this experience. Our participants included 144 news readers, equally distributed among three groups: Chinese immigrants, Vietnamese immigrants, and local residents living in the same region. We analyzed Q&A logs generated from each participant's interaction with the chatbot during the news reading process. We also collected the takeaways participants derived from the news reading through a retelling task. Seven distinct categories of questions were identified from our analysis of the Q&A logs. Compared to the local group, both immigrant groups spent more effort understanding the literal meaning and practical implications of the news content, yet engaged less in analytical thinking to assess the news content. Furthermore, both immigrant groups relied more heavily on information provided by the chatbot's responses when formulating their takeaways regarding actions to take in life. We discussed our findings in the context of chatbot design for a reader-centered news experience. We argued that chatbots should be built to support immigrant news readers in diversifying their sensemaking perspectives while also being cautious about relying on any single information source.

## ACKNOWLEDGMENTS

This research was supported by the National Science Foundation, under grants #2147292 and #2229885. We thank Jimmy Dang, Henry Tran, Tien Le, Yimin Xiao, Victoria Chang, Yuhang Zhou, and Cartor Hancock for their assistance. We thank all of our participants for their time and effort, which made this work possible. We also thank the anonymous reviewers for their valuable feedback on earlier versions of this paper.

## APPENDIX A

| | News Article 1 | News Article 2 | News Article 3 | News Article 4 |
|---|---|---|---|---|
| Word count | 546 | 762 | 500 | 692 |
| I find the information provided for the topic discussed in this news article is: Limited - Comprehensive | 5.13/1.25 | 5.33/1.41 | 5.75/1.28 | 5.33/1.00 |
| I find the information provided for the topic discussed in this news article is: Straightforward - Complex | 3.63/1.41 | 3.44/1.59 | 4.50/1.77 | 3.78/1.79 |
| I find the information provided for the topic discussed in this news article is: Shallow - In-depth | 5.13/1.73 | 4.89/0.93 | 5.75/1.28 | 5.22/0.67 |
| While reading this news article, I feel its content is: Ambiguous - Clear | 6.13/0.64 | 5.78/0.67 | 5.50/1.60 | 5.22/1.79 |
| While reading this news article, I feel its content is: Disorganized - Logically structured | 5.75/1.28 | 5.00/1.41 | 5.38/1.60 | 5.00/2.12 |
| While reading this news article, I feel its content is: Confusing - Accessible | 5.63/1.19 | 5.44/1.13 | 4.75/1.67 | 5.11/1.62 |
| The voice or perspective in this news article appears: Causal - Professional | 5.38/1.30 | 5.67/1.32 | 5.88/1.36 | 6.00/1.00 |
| The voice or perspective in this news article appears: Superficial - Insightful | 5.88/0.83 | 5.44/1.13 | 5.25/1.39 | 5.56/0.88 |
| The voice or perspective in this news article appears: Unsubstantiated - Validated | 5.75/1.04 | 5.22/1.20 | 5.63/1.19 | 5.56/1.01 |
| I find the overall tone of this news article is: Negative - Positive | 3.63/0.92 | 4.67/1.50 | 3.63/0.74 | 4.67/1.32 |
| I find the overall tone of this news article is: Subjective - Objective | 5.50/1.31 | 4.67/1.50 | 5.38/0.92 | 5.22/1.72 |
| I find the overall tone of this news article is: Informal - Formal | 5.50/1.07 | 5.00/1.73 | 5.50/1.20 | 5.67/1.00 |

**Note:* This table summarizes the evaluations of the four news articles utilized in our study. For each news article, we present its word count along with subjective ratings provided by Prolific workers on 7-point Likert scales. These ratings encompass several dimensions, including comprehensiveness, depth, clarity, structure, and tone of the content, as well as the style and perceived professionalism of the writing. Each cell displays the mean and standard deviation of the corresponding evaluation.